\documentclass[aps,prl,twocolumn,amsmath,amssymb,showpacs]{revtex4}

\usepackage{graphicx}

\begin{document}

\title{%
Emergent Lorentz Symmetry with Vanishing Velocity in a Critical
Two-Subband Quantum Wire
}

\author{%
M.~Sitte$^1$, A.~Rosch$^1$, J.~S.~Meyer$^2$, K.~A.~Matveev$^3$, and
M.~Garst$^1$
}

\affiliation{%
$^1$Institut f\"ur Theoretische Physik, Universit\"at zu K\"oln,
Z\"ulpicher Stra\ss e.~77, 50937 K\"oln, Germany \\
$^2$Department of Physics, The Ohio State University, Columbus, Ohio
43210, USA \\
$^3$Materials Science Division, Argonne National Laboratory, Argonne,
Illinois 60439, USA
}

\date{%
\today
}

\begin{abstract}
We consider a quantum wire with two subbands of spin-polarized electrons
in the presence of strong interactions. We focus on the quantum phase
transition when the second subband starts to get filled as a function of
gate voltage. Performing a one-loop renormalization group analysis of
the effective Hamiltonian, we identify the critical fixed-point theory
as a conformal field theory having an enhanced SU(2) symmetry and
central charge $3/2$. While the fixed point is Lorentz invariant, the
effective ``speed of light'' nevertheless vanishes at low energies due
to marginally irrelevant operators leading to a  diverging critical
specific heat coefficient.
\end{abstract}


\maketitle

As a function of an external gate voltage, the conductance of quantum
wires increases in steps of height $G_0 = 2 e^2/h$ when more and more
subbands are subsequently filled. Each conductance plateau transition
defines a quantum critical point where the ground state changes its
character. Close to each critical point one expects universal scaling
and a rich behavior as a function of gate voltage and temperature. These
quantum phase transitions and especially their transport signatures are,
however, only partially understood \cite{Balents00, Meyer07, Meyer08,
Garst08}.

A quantum phase transition is characterized by the critical degrees of
freedom, their dynamics, and the underlying symmetries at the critical
point \cite{review}. In one-dimensional systems, many quantum critical
points are Lorentz invariant, and conformal field theory (CFT) provides
a valuable scheme to classify them. In such a case, the powerful
machinery available for CFTs allows the calculation of the critical
behavior and the systematic investigation of the leading corrections to
scaling. In particular, a characteristic of CFTs is a linear specific
heat, $C = c (\pi/3v)T$, inversely proportional to the velocity $v$ of
excitations, that identifies the central charge $c$ measuring the
effective number of degrees of freedom \cite{Bloete86, Affleck86}.

In this Letter we study a model of spinless fermions close to the
quantum critical point where a second subband is about to get filled as
a function of gate voltage. Three different phases \cite{Meyer07,
Meyer08} are relevant to describe this system. First, for a small
chemical potential, only the first subband is occupied, and the electron
system forms an ordinary Luttinger liquid. Second, for a large chemical
potential, both subbands are filled, and one obtains two Luttinger
liquids. There is, however, also a third phase  which can be visualized
as a (fluctuating) zigzag Wigner crystal, where the relative motion of
charges in the two subbands is locked and only one gapless mode
survives.

Upon increasing the chemical potential, the one-dimensional phase
undergoes a transition to the fluctuating zigzag Wigner crystal phase
\cite{Meyer07, Meyer08}. The nature of this transition depends on the
interaction strength. For weak interactions, it has been argued
\cite{Balents00, Meyer07} that thermodynamic properties at the critical
point, where the first phase becomes unstable, are not qualitatively
changed by the interactions (see below). One obtains a Lifshitz
transition describing the filling of a noninteracting subband of
fermions with a quadratic dispersion. More precisely, the interactions
turn out to be dangerously irrelevant. While they do not affect the
properties at criticality, they nevertheless induce a zigzag Wigner
crystal for small but finite filling of the second subband. When the
interactions are sufficiently strong, however, the universality class of
the transition changes. The identification of the critical properties at
such strong interactions is the aim of this Letter.

We show that the corresponding transition has a rather unusual
structure. On the one hand, we find that the critical fixed point is
given by a well-known CFT with central charge $c = 3/2$, describing
three Majorana fermions with equal velocity. As in many other systems
the symmetries at the critical point are enhanced as neither Lorentz
invariance nor the SU(2) symmetry of the $c = 3/2$ CFT are properties of
the original model. On the other hand, it turns out that, at
criticality, the effective velocity---the ``speed of light'' of the
CFT---vanishes in the low-energy limit. This peculiar behavior arises
from marginally irrelevant corrections to the CFT that spoil Lorentz
invariance. It is thus not a property of the fixed point itself, but a
consequence of the slow flow towards this fixed point. We discuss the
conditions under which a vanishing of the critical velocity occurs
generically. A direct consequence of this behavior is a divergent
critical specific heat coefficient, $\gamma = C/T$, for $T \to 0$.

A known example of a similar situation occurs in a model for
two-dimensional electrons in graphene interacting via a long-range
Coulomb potential \cite{Abrikosov70, Gonzalez}. The Fermi points in
graphene give rise to an emergent Lorentz invariance that is, however,
broken by the static Coulomb interaction. Because of the vanishing
density of states at the Fermi points, the interaction remains
unscreened and thus long-range. Although the effective interaction is
found to be marginally irrelevant, Lorentz invariance remains broken in
the low-energy limit. In contrast to our case, the effective velocity is
logarithmically enhanced in the presence of Coulomb interaction,
resulting in a suppression of the specific heat.

\begin{figure}[b]
\includegraphics[width=0.19\textwidth]{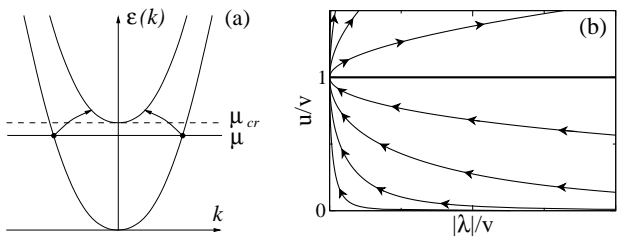}
\quad
\includegraphics[width=0.24\textwidth]{fig2}
\caption{%
(a) Pair-tunneling process between the two subbands. The chemical
potential $\mu$ is close to the critical value $\mu_{\rm cr}$ ($r \equiv
\mu_{\rm cr} - \mu$). (b) One-loop
RG flow diagram of the effective model $\mathcal{L}_{\rm eff}$
determined by the invariant $\mathcal{I}=\lambda^2 K v u/(u-v)^4$ [see
Eqs.~(\ref{RGa}) and (\ref{RGb})]. $\lambda$ is the interaction, $u$ is
the pair tunneling and $v$ is the plasmon velocity. There is a line of
fixed points at $u = v$ parametrized by the dynamical exponent
(\ref{zExponent}).
}
\label{fig:PairTunneling&RGflow}
\end{figure}

To be specific, we study the model Hamiltonian $\mathcal{H} =
\mathcal{H}_{1} + \mathcal{H}_{2} + \mathcal{H}_{12}$, where
\cite{Meyer07}
\begin{gather}
\label{Band1}
\mathcal{H}_{1} = \frac{v}{2\pi} \int dx \biggl[ K (\partial_x \theta)^2
+ \frac{1}{K} (\partial_x \phi)^2 \biggr], \\
\label{Band2}
\mathcal{H}_{2} = \int dx \, \psi^\dagger \biggl(
-\frac{\partial_x^2}{2m} - \mu + \mu_{\rm cr} \biggr) \psi, \\
\label{interaction}
\mathcal{H}_{\rm 12} = \int dx \biggl[ -\frac{g_x}{\pi} \partial_x \phi
\, \psi^\dagger \psi + \frac{u}{2} (e^{i 2 \theta} \, \psi \partial_x
\psi  + {\rm H.c.}) \biggr].
\end{gather}
The first subband [see Fig.~\ref{fig:PairTunneling&RGflow}(a)] has a
finite filling and can be treated as a Luttinger liquid
[Eq.~(\ref{Band1})]. Its excitations are plasmons, propagating waves of
electron density, $\delta n = -\frac{1}{\pi} \partial_x \phi$,
represented by the bosonic field $\phi$, and $\theta$ is its conjugate
with $[\frac{1}{\pi} \partial_x \phi(x), \theta(x')] = -i \delta(x-x')$.
The second subband, on the other hand, contains spinless
non-relativistic fermions [Eq.~(\ref{Band2})] that are close to the
quantum phase transition occurring at $r \equiv \mu_{\rm cr} - \mu = 0$
when $\mu$ is increased (e.g., by tuning a gate voltage). Near this
transition, the interaction among fermions in the second subband is
negligible due to the Pauli principle. The Galilean invariance of
$\mathcal{H}_2$ is manifest and implies a critical dynamical exponent
$z = 2$ for the fermions in the absence of intersubband interactions.
However, there are two important types of such interactions. The first
term in Eq.~(\ref{interaction}) is the density-density interaction,
$g_x$, and the second term, $u$, arises from pair-tunneling processes of
electrons  as depicted in Fig.~\ref{fig:PairTunneling&RGflow}(a). We
assume $u\geq0$. Interestingly, the nature of the transition at $r = 0$
was found to depend on the interaction strength \cite{Meyer07}.

First, consider the Hamiltonian $\mathcal{H}$ without pair tunneling,
$u = 0$. At criticality, $r=0$, there is a van Hove singularity in the
tunneling density of states (DOS) that results in a strong hybridization
between the modes of the two subbands in the presence of a finite
interaction, $g_x$. At their band bottom, the fermions are very slow,
and the fast plasmons can adiabatically adjust themselves to the
fermions in the second subband. Explicit calculations \cite{Balents00}
show that the coupling of the resulting dressed fermions to the
Luttinger liquid is irrelevant at the critical point. Therefore the
transition remains a usual one-dimensional Lifshitz transition. For
$\mu < \mu_{\rm cr}$ the chemical potential is below the band edge,
implying a charge gap $\Delta = \mu_{\rm cr} - \mu = r$. In the limit of
a weak interaction $g_x$, this scenario remains valid even in the
presence of a finite pair tunneling $u$. However, it was observed in
Ref.~\cite{Meyer07}, that, for $r<0$, the pair tunneling $u$ leads to a
ground state with a zigzag charge density wave associated with a gap,
$\Delta \sim (-r)^\alpha$, where the exponent is large, $\alpha \gg 1$,
in the weak coupling limit. As a consequence, the actual conductance
plateau transition does not coincide with the Lifshitz transition at
$r = 0$, but is rather shifted to a negative value of $r < 0$, i.e., a
finite filling of the second subband \cite{Meyer08}.

The universality class of the transition at $r = 0$ changes
qualitatively when the interaction $g_x$ increases.  To understand the
nature of this transition, we apply a unitary transformation, $U =
e^{i \int \theta \psi^\dagger \psi \, dx}$, to the Hamiltonian such that
effectively $\psi \to U^\dagger \psi U = \psi e^{-i \theta}$ and
$\partial_x \phi \to \partial_x \phi + \pi \psi^\dagger \psi$. Switching
to a Euclidean Lagrangian description, the effective model then can be
written as $\mathcal{L}_{\rm eff} = \mathcal{L}_{\rm LL} +
\mathcal{L}_{\rm Ising} + \mathcal{L}_{\rm int}$, where
\begin{gather}
\label{LLL}
\mathcal{L}_{\rm LL} = \frac{1}{2 \pi v K} [(\partial_\tau \phi)^2 + v^2
(\partial_x\phi)^2], \\
\label{LIsing}
\mathcal{L}_{\rm Ising} = \psi^\dagger \partial_\tau \psi  + \frac{u}{2}
(\psi \partial_x \psi + {\rm H.c.}) + r \, \psi^\dagger \psi, \\
\label{LInteraction}
\mathcal{L}_{\rm int} = -\lambda \frac{1}{\pi} \partial_x\phi \,
\psi^\dagger \psi,
\end{gather}
$\tau$ is the imaginary time, and $\lambda \equiv g_x - \pi v/K$.

For $\lambda = 0$, the plasmonic and fermionic sectors decouple, but the
fermionic part $\mathcal{L}_{\rm Ising}$ can now be identified with an
Ising model that is critical for $r = 0$. The term $\psi^\dagger
\partial_x^2 \psi$ in Eq.~(\ref{Band2}) is negligible close to
criticality as it is subleading compared with the spatial gradient term
in $\mathcal{L}_{\rm Ising}$. As a consequence, an effective Lorentz
invariance for the critical fermionic excitations emerges, leading to a
dynamical exponent $z = 1$. Note, however, that there are two different
velocities: the fermions propagate with velocity $u$ and the plasmons
with velocity $v$. Moreover, the effective model now possesses the
self-duality symmetry of the Ising model which guarantees that, for
$\lambda = 0$, there is a gap $\Delta = |r|$ in the fermionic sector on
either side of the transition.

In the present work, we analyze the effective model
$\mathcal{L}_{\rm eff}$ perturbatively in the residual interaction
$\lambda$ \cite{remark}, which explicitly breaks the Lorentz invariance
of both the plasmonic and the fermionic excitations. We show that this
Lorentz-invariance breaking leaves traces in the critical behavior
although the interaction $\lambda$ is found to be (marginally)
irrelevant in the renormalization group (RG) sense.

\begin{figure}[b]
\includegraphics[width=0.4\textwidth]{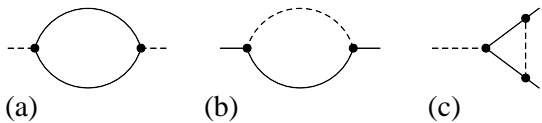}
\caption{%
One-loop perturbative corrections. The dashed and solid lines represent
the plasmon and fermion propagator, respectively. The dot is the
interaction vertex $\lambda$ given in Eq.~(\ref{LInteraction}). (a)
Plasmon self-energy, (b) fermion self-energy, and (c) vertex-correction.
}
\label{fig:diagrams}
\end{figure}

Perturbation theory in $\lambda$ is accompanied by logarithmic IR
singularities that can be absorbed into a renormalization group flow of
the parameters of the model. Introducing a momentum cutoff $\Lambda$, we
perform a Wilsonian renormalization group step by integrating out
perturbatively modes within a momentum shell $(\Lambda/b, \Lambda)$ with
$b>1$, and we calculate the plasmon and fermion self-energy and the
vertex correction to one-loop order (see Fig.~\ref{fig:diagrams}). After
rescaling momenta $k \to k/b$ and frequencies $\omega \to \omega/b^z$,
where $z$ is a so-far-unspecified dynamical exponent, the momentum
cutoff of the perturbatively corrected theory is restored to $\Lambda$.
Moreover, introducing wave function renormalizations, $\phi \to
\sqrt{Z_B}\, \phi$ and $\psi \to \sqrt{Z_F}\, \psi$, we can choose the
RG conditions such that the bosonic velocity $v$, the Luttinger
parameter $K$, and the unit prefactor of $\psi^\dagger \partial_\tau
\psi$ in (\ref{LIsing}) are RG invariants. This can be achieved by
introducing an anomalous dynamical exponent
\begin{equation}
\label{zExponent}
z = 1 + \frac{\lambda^2 K}{4 \pi^2 u v}.
\end{equation}
A dynamical exponent $z \not= 1$ is a manifestation of the broken
Lorentz invariance. Note that within this RG scheme the velocities $u$
and $v$ are not the physical velocities. In particular, the plasmon
velocity $v_{\rm phys}$ flows under RG (see below).

The fermionic velocity $u$, the interaction $\lambda$, and the effective
chemical potential $r$ obey the one-loop RG equations:
\begin{gather}
\label{RGa}
\frac{\partial u}{\partial \ln b} = -\frac{\lambda^2 K}{\pi^2} \biggl[
\frac{1}{(u + v)^2} - \frac{1}{4 u v} \biggr] u, \\
\label{RGb}
\frac{\partial \lambda}{\partial \ln b} = -\frac{\lambda^3 K}{2 \pi^2}
\biggl[ \frac{1}{u (v + u)} + \frac{1}{(u + v)^2} - \frac{3}{4 u v}
\biggr], \\
\label{RGc}
\frac{\partial  r}{\partial \ln b} = r - \frac{r \lambda^2 K}{2 \pi^2}
\biggl[ \frac{1}{u (v + u)} + \frac{1}{(u + v)^2} - \frac{1}{2 u v}
\biggr].
\end{gather}
These coupled nonlinear differential equations can be solved
analytically by realizing that the quantity $\mathcal{I} = \lambda^2 K v
u/(u-v)^4$ is an invariant of the RG flow. $\mathcal{I}$ determines
directly the flow diagram shown in
Fig.~\ref{fig:PairTunneling&RGflow}(b). If the fermionic velocity is
larger than the bosonic velocity, $u > v$, the RG flow is towards strong
coupling, $\lambda \to \infty$, which may indicate either a first order
transition or a different strong coupling fixed point. If the fermionic
and bosonic velocities coincide, $u = v$, the one-loop beta functions
(\ref{RGa}) and (\ref{RGb}) vanish, and one obtains a line of fixed
points parametrized by the dynamical exponent $z$
[Eq.~(\ref{zExponent})]. This line of fixed points is, however, likely
to be an artifact of the one-loop RG. Fortunately, in a two-subband
quantum wire we expect the pair-tunneling $u$ to be much smaller than
the plasmon velocity of the first subband $v$ \cite{remark}. In this
regime, the RG flow is towards weak coupling, $\lambda \to 0$, and is
fully controlled by our one-loop analysis \cite{remark2}.

If the flow is towards weak coupling, the fermionic velocity $u(b)$
approaches the bosonic velocity $v$ in the limit $b\to \infty$ [see
Fig.~\ref{fig:PairTunneling&RGflow}(b)]. As a consequence, the
weak-coupling fixed point is characterized by a single velocity.
Diagonalizing the fermionic sector and refermionizing the bosonic mode
of the effective model $\mathcal{L}_{\rm eff}$ for $\lambda = 0$ and
$u = v$, the fixed-point Hamiltonian can be expressed in terms of three
left- and right-moving Majorana fields, $L_a$ and $R_a$, respectively:
\begin{equation}
\label{FP}
\mathcal{H}_{\rm FP} = \frac{i}{2} \sum_{a=1}^3 v [L_a \partial_x L_a -
R_a \partial_x R_a].
\end{equation}
At the critical point one has an enhanced SU(2) symmetry describing a
rotation in the space of the three real Majorana fields
\cite{Tsvelik90}. Furthermore, the fixed point is Lorentz invariant.
Nevertheless, we show below that the residual breaking of Lorentz
invariance due to the interaction $\lambda$ leads to important
corrections to the fixed-point Hamiltonian (\ref{FP}), although
$\lambda$ is formally irrelevant.

The flow towards the fixed point can be determined by considering the
limit of a  small velocity difference: $w \equiv 1 - u/v \ll 1$. By
expanding the RG equations (\ref{RGa}) and (\ref{RGb}) in lowest order
in $w$, one finds
\begin{equation}
\label{RGEqu2}
\frac{\partial w}{\partial \ln b} = -\frac{K}{16 \pi^2 v^2 } \lambda^2
w^2, \quad \frac{\partial \lambda}{\partial \ln b} =
-\frac{K}{8 \pi^2 v^2 } \lambda^3 w.
\end{equation}
In this regime, $u$ approaches $v$ very slowly,
\begin{equation}
\label{running-u1}
u(b) \approx  v \biggl( 1 - \frac{\mathcal{A}}{(\ln b)^{1/5}} \biggr),
\end{equation}
where the prefactor $\mathcal{A} = (16 \pi^2/5 \, \mathcal{I})^{1/5}$
has been obtained from the exact solution of (\ref{RGa}) and
(\ref{RGb}).

If the pair tunneling is very small, $u \ll v$, there is an extended
regime where the initial flow differs qualitatively from the asymptotic
behavior (\ref{RGEqu2}):
\begin{equation}
\label{running-u2}
u(b) \approx \sqrt{u^2 + v^2 \frac{\mathcal{I}}{2 \pi^2} \ln b} \;.
\end{equation}
An estimate of the crossover scale $b^*$ between the two regimes yields
$\ln b^*\sim 1/\mathcal{I}$. This scale is very large in the limit $u
\ll v$ such that the asymptotic behavior is attained only at extremely
low energies.

\begin{figure}[t]
\includegraphics[width=0.4\textwidth]{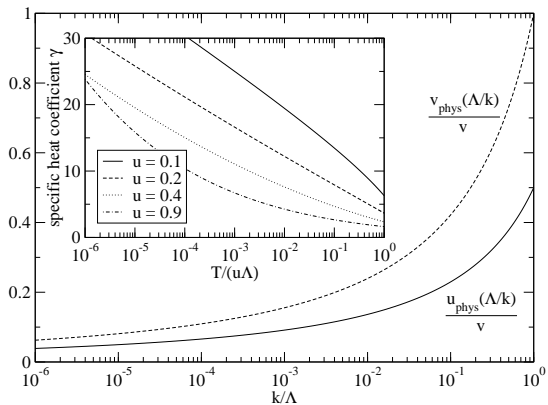}
\caption{%
Momentum dependence of the physical velocities of the Luttinger liquid,
$v_{\rm phys}(\Lambda/k)$, and the excitations of the second subband,
$u_{\rm phys}(\Lambda/k)$, for $K=1$, $|\lambda|=v=1$ and $u=0.5$ at the
quantum critical point. For $k \to 0$ both velocities vanish while
$u_{\rm phys}(\Lambda/k)/v_{\rm phys}(\Lambda/k) \to 1$. Inset: specific
heat coefficient for various values of $u$.
}
\label{fig:u-flow}
\end{figure}

When interpreting the results given above, it is essential to realize
that they have been obtained in an RG scheme where time and space have
been rescaled in an asymmetric way as $z\neq 1$. As velocity is length
over time, this asymmetric scaling modifies the physical velocities by
an extra factor
\begin{equation}
\label{vcorr}
\Omega(b) = \exp \biggl[ \int_0^{\ln b} (1-z(b')) \, d \ln b' \biggr]
\end{equation}
with $v_{\rm phys} = \Omega \, v$, $u_{\rm phys} = \Omega\, u$. For low
energies, i.e., $\ln b \to \infty$, $z(b)$ approaches unity because the
fixed point is Lorentz invariant:
\begin{equation}
\label{zcorr}
z(b) \approx 1 + \frac{2}{5} \biggl( \frac{10 \, \mathcal{I}}{\pi^2}
\biggr)^{1/5} \frac{1}{(\ln b)^{4/5}}.
\end{equation}
However, the flow towards $z = 1$ is so slow that the integral $\int
[z(b')-1] \, d \ln b'$ in (\ref{vcorr}) diverges and, thus,
$\Omega(b \to \infty) = 0$. This implies that at the fixed point,
$\ln b=\infty$,  all physical velocities vanish. In
Fig.~\ref{fig:u-flow} the momentum dependence of the two physical
velocities $u_{\rm phys}$ and $v_{\rm phys}$ is shown. As described by
Eq.~(\ref{running-u1}) the two velocities approach each other slowly,
and both vanish for $k\to 0$ due to the $z$ renormalization of
Eq.~(\ref{vcorr}).

The vanishing of the velocities naturally leads to a divergence of the
specific heat coefficient $\gamma$. At criticality ($r=0$), $\gamma$
scales as the inverse of the physical velocities, $\gamma(b) =
\frac{\pi}{6 \Omega(b)} \bigl( \frac{2}{v} + \frac{1}{u(b)} \bigr)$. The
RG flow is stopped when $T(b)/u(b)$ reaches the cutoff, where
temperature obeys $\partial T/\partial \ln b = z T$. From
Eqs.~(\ref{vcorr}) and (\ref{zcorr}), one then obtains asymptotically
\begin{equation}
\label{gamma}
\gamma \sim \exp \biggl[ 2 \biggl( \frac{10 \, \mathcal{I}}{\pi^2} \,
\ln \frac{v \Lambda}{T} \biggr)^{1/5} \biggr].
\end{equation}
The inset of Fig.~\ref{fig:u-flow} shows the $T$ dependence of $\gamma$.

Away from criticality, $r\neq 0$, the fermionic excitations are gapped.
The gap $\Delta$ can be obtained by integrating the RG equations
(\ref{RGa})--(\ref{RGc}) up to $b = b_\Delta$ with $r(b_\Delta) =
u(b_\Delta) \Lambda$. In physical units, this corresponds to the energy
scale $\Delta = u(b_\Delta) \Omega(b_\Delta) \Lambda/b_\Delta$, where we
have again taken into account the rescaling of the time axis. Compared
to the Ising relation, $\Delta = |r|$, the resulting $T = 0$ gap,
\begin{equation}
\label{gap}
\Delta \sim |r|\, \exp \biggl[ -3 \biggl( \frac{10 \,
\mathcal{I}}{\pi^2} \ln \frac{v \Lambda}{|r|} \biggr)^{1/5} \biggr],
\end{equation}
is suppressed due to the residual interaction $\lambda$. The suppression
is, however, slower than power law. The gap $\Delta$ should also be
observable in the tunneling DOS.

In this Letter we have shown that the quantum phase transition from a
Luttinger liquid to a zigzag Wigner crystal has a fixed point given by a Lorentz-invariant theory of three noninteracting Majorana fermions.
However, the effective velocities of the modes vanish for low energies.
This behavior will occur quite generally for systems where marginally
irrelevant terms break the Lorentz invariance. More precisely, whenever
the correction to the dynamical critical exponent vanishes with
$1/\ln b$ or slower, the integral in (\ref{vcorr}) diverges such
that---depending on the sign of $(z-1)$---the effective velocity will
either diverge or vanish for low energies.

We thank N. Andrei, L. Balents, T. Senthil,  and M. Vojta for useful
discussions. This work was supported by the DFG through SFB~608 and by
the U. S. Department of Energy, Office of Science, under Contract
No.~DE-AC02-06CH11357 and No.~DE-FG02-07ER46424.


\begin{thebibliography}{10}
\bibitem{remark}
For quantum wires with Coulomb interaction screened by a gate at
distance $d$, the parameters $g_x$, $v$, and $K$ are related
\cite{Meyer07}, yielding $\lambda \simeq -\pi v K = -\pi v_{F1}$. The
perturbative RG in $\lambda$ is controlled by the small parameter
$\lambda^2 K/(v u) \sim v^2_{F1}/[e^4 \log(k_{F1} d)]$. Here $v_{F1}$
and $k_{F1}$ is the Fermi velocity and momentum of the filled band,
respectively, and $\log (k_{F1} d) \gg 1$. Note that the same estimates
give $(u/v)^2 \sim e^2/[v_{F1} \log (k_{F1} d )] \ll 1$ for the ratio of
velocities.
\bibitem{remark2}
The one-loop RG is controlled at least as long as $v - u > \sqrt{K}
|\lambda|$, which follows from the worst-case assumption that the
two-loop corrections to the RG equations (\ref{RGEqu2}) are of order $\mathcal{O}((v-u)^0)$.
\end{thebibliography}
\end{document}